\documentclass[preprint,epsfig,floats,aps]{revtex4}
\usepackage{epsfig}
\begin{document}
\title{Energy and centrality dependences of charged multiplicity
       density in relativistic nuclear collisions}
\author{Ben-Hao Sa$^{1,5,6}$\footnotemark,
 \footnotetext{E-mail: sabh@iris.ciae.ac.cn}
 A. Bonasera$^2$\footnotemark,
 \footnotetext{E-mail: bonasera@lns.infn.it}
 An Tai$^3$, and Dai-Mei Zhou$^4$}
\affiliation{
$^1$  China Institute of Atomic Energy, P. O. Box 275 (18),
      Beijing, 102413 China \\
$^2$  Laboratorio Nazionale del Sud, Instituto Nazionale Di Fisica Nucleare,
      v. S. Sofia 44 95132 Catania, Italy \\
$^3$  Department of Physics and Astronomy, University of California,
      at Los Angeles, Los Angeles, CA 90095 USA \\
$^4$  Institute of Particle Physics, Huazhong Normal University,
      Wuhan, 430079 China\\
$^5$  CCAST (World Lab.), P. O. Box 8730 Beijing, 100080 China\\
$^6$  Institute of Theoretical Physics, Academia Sinica, Beijing,
      100080 China
}
\begin{abstract}
Using a hadron and string cascade model, JPCIAE, the energy and centrality
dependences of charged particle pseudorapidity density in relativistic
nuclear collisions were studied. Within the framework of this model, both the 
relativistic $p+\bar p$ experimental data and the PHOBOS and PHENIX $Au+Au$ 
data at $\sqrt s_{nn}$=130 GeV could be reproduced fairly well without retuning 
the model parameters. The predictions for full RHIC energy $Au+Au$ collisions 
and for $Pb+Pb$ collisions at the ALICE energy were given. Participant nucleon 
distributions were calculated based on different methods. It was found that
the number of participant nucleons, $<N_{part}>$, is not a well defined 
variable both experimentally and theoretically. Therefore, it is inappropriate 
to use charged particle pseudorapidity density per participant pair as a 
function of $<N_{part}>$ for distinguishing various theoretical models.
\end{abstract}
\maketitle

The main focus of the Relativistic Heavy-Ion Collider (RHIC) at Brookhaven 
National Laboratory (BNL) is to explore the phase transition related to 
the quark deconfinement and the chiral symmetry restoration. The first 
available experimental data were the energy dependence of charged particle 
pseudorapidity density in $Au+Au$ collisions at $\sqrt s_{nn}$=56 and 130 GeV 
from the PHOBOS collaboration \cite{pho1}. After that, the PHENIX collaboration 
published their data of centrality dependence of the charged particle 
pseudorapidity density in $Au+Au$ collisions at $\sqrt s_{nn}$=130 GeV 
\cite{phe1}.

\begin{figure}[ht]
\centerline{\hspace{-0.5in}
\epsfig{file=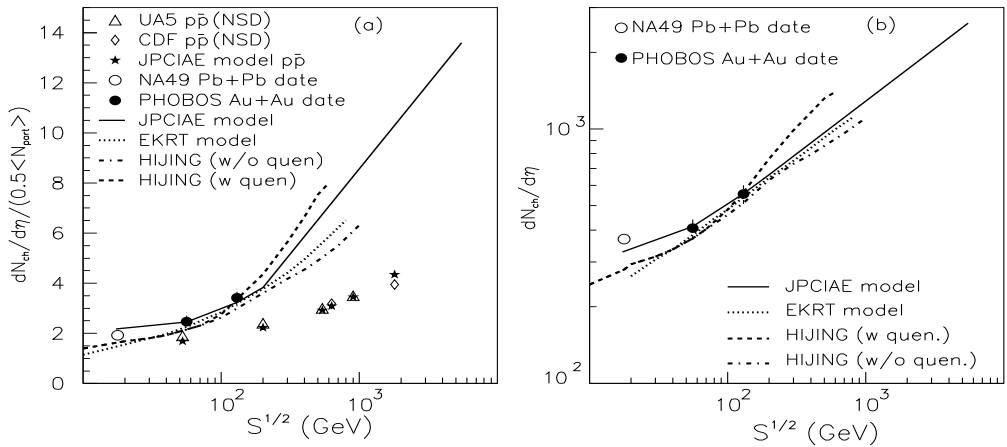,width=5.0in,height=2.3in,angle=0}}
\caption{The energy dependence of the charged particle pseudorapidity density 
at mid-pseudorapidity in relativistic $p+\bar p$ and $A+A$ collisions.}
\label{fig1}
\end{figure}

 It has been predicted that the rare high charged multiplicity could indicate 
the onset of the Quark-Gluon-Plasma (QGP) phase, since the extra entropy in 
the QGP phase could manifest itself as a huge number of produced particles in 
the final state \cite{hov,gor,kap1}. On the other hand, in \cite{wang1}  
the centrality dependence of charged multiplicity has been proposed to  
provide information on the relative importance of soft versus hard 
processes in particle production and therefore provide a means of 
distinguishing various theoretical models for particle production. 

 The pQCD calculation with assumption of gluon saturation \cite{esk1} (referred 
to as EKRT model later) was first used to study the centrality dependence 
of the charged particle pseudorapidity density at RHIC. In \cite{wang1} the 
HIJING model met with success in describing both the energy and centrality 
dependence of the charged particle pseudorapidity density. The conventional 
eikonal approach and the high density QCD (referred to as KN model later) 
\cite{kha1} were also used to investigate the centrality dependence and the 
both methods surprisingly obtained almost identical centrality dependence. 
Recently, authors in \cite{cape} reported their results from the Dual Parton 
Model. It was found that the experimental observation, the charged particle 
pseudorapidity density per participant pair slightly increasing with 
$<N_{part}>$, was reproduced by \cite{wang1,kha1,cape}, but contradicted the 
results of \cite{esk1}.   

 In this letter a hadron and string cascade model, JPCIAE \cite{sa1}, was
employed to study this issue further. Within the framework of this model 
the experimentally measured energy dependence of the charged particle  
mid-pseudorapidity density per participant pair both in relativistic $p+\bar p$ 
and $Au+Au$ collisions at RHIC was reproduced fairly well without retuning the 
model parameters. The predictions for the full RHIC energy $Au+Au$ collisions 
and for $Pb+Pb$ collisions at the ALICE energy were also given. In studying 
centrality dependence the focus was put on the calculations of $<N_{part}>$, 
its definition and uncertainty. Both the PHENIX \cite{phe1} and the PHOBOS 
\cite{pho2} observations that the charged particle mid-pseudorapidity density 
per participant pair slightly increases with $<N_{part}>$ could be reproduced 
fairly well by JPCIAE. However, this study indicated that it is not suitable to use the charged particle mid-pseudorapidity density per participant pair as a 
function of $<N_{part}>$ to constrain theoretical models for particle 
production, because $<N_{part}>$ is not a well defined physical variable both 
experimentally and theoretically.  

 The JPCIAE model was developed based on PYTHIA \cite{sjo1}. In the JPCIAE 
model the nucleons in a colliding nucleus are distributed randomly in the 
sphere of the nucleus with a radius of 1.12$A^{1/3}$ fm. The modules of the 
nucleons are sampled by the Woods-Saxon distribution and the solid angles of 
the nucleons are sampled uniformly in 4$\pi$. Each nucleon is given a  beam 
momentum in z direction and zero initial momentum in x and y directions. After 
the construction of initial particle list the collision time of each colliding 
pair is calculated under the requirement that the least approaching distance of the colliding pair along their Newton straight-line trajectory should be 
smaller than $\displaystyle{\sqrt{\sigma_{tot}/\pi}}$, where $\sigma_{tot}$ 
refers to the total cross section. The nucleon-nucleon collision with the 
least collision time is then selected from the initial collision list to 
perform the first collision. After the first collision, both the particle list 
and the collision list are updated and now the collision list may consist 
of not only nucleon-nucleon collisions, but also collisions between produced 
particles and the nucleons and between produced particles themselves. The next 
collision is selected from the new collision list. The processes proceed until
the collision list is empty. 

For each collision pair, if its CMS energy is larger than a given cut,
we assume that strings are formed after the collision and PYTHIA is used
to deal with particle production. Otherwise, the collision is treated as a
two-body collision \cite{cugn,bert,tai1}. The cut (=4 GeV in the program) 
was chosen by observing that JPCIAE correctly reproduces charged multiplicity 
distributions in AA collisions. 

It should be noted here that the JPCIAE model is not a simple superposition of 
nucleon-nucleon collisions since the rescatterings among participant, spectator 
nucleons and produced particles are taken into account. We refer to \cite{sa1} 
for more details of the JPCIAE model.    

\begin{figure}[ht]
\centerline{\hspace{-0.5in}
\epsfig{file=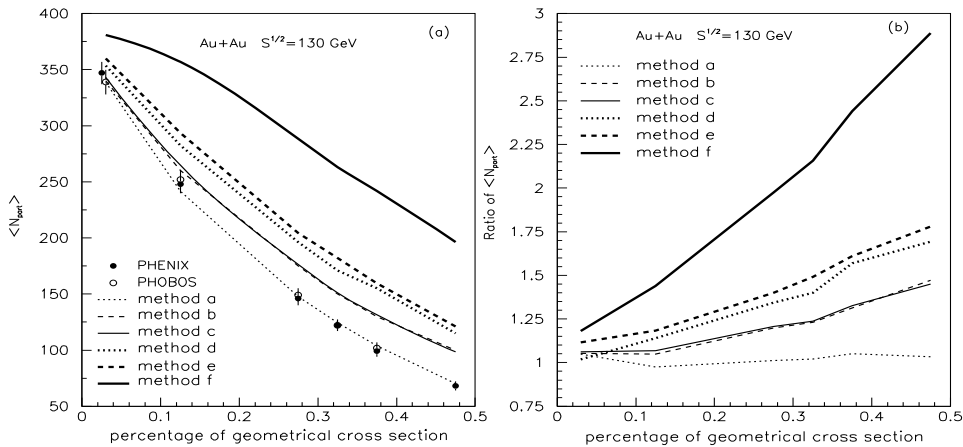,width=5.0in,height=2.3in,angle=0}}
\caption{The number of participant nucleons $<N_{part}>$, (a), and the ratios 
of the different curves in (a) to the PHENIX data, (b), as a function of the 
percentage of geometrical cross section.}
\label{fig2}
\end{figure}

 In Fig.1(a) the experimental data of charged particle pseudorapidity 
density per participant pair at mid-pseudorapidity in relativistic $p+\bar p$ 
(open triangles and rhombuses with error bar) and $A+A$ collisions (open 
circles with error bar for $Pb+Pb$ at SPS and full circles with error 
bar for $Au+Au$ at RHIC) \cite{pho1} are compared with JPCIAE model
(full stars for $p+\bar p$ and the solid curve for $A+A$ collisions), HIJING 
model (dotted-dash curve without jet quenching and dashed curve with jet 
quenching) \cite{wang1} and EKRT model (dotted curve) \cite{esk1}. The data of 
both $p+\bar p$ and $A+A$ collisions at relativistic energies were reproduced 
fairly well by JPCIAE model without retuning model parameters.  Fig. 1 (b) is 
the same as (a) but the vertical coordinate here is the charged particle 
pseudorapidity density itself. The JPCIAE model predictions for full RHIC 
energy $Au+Au$ collisions and for $Pb+Pb$ collisions at the ALICE energy in 
both panels may supply a benchmark for QGP formation since the QGP phase is 
not included in the JPCIAE model.  

 Since number of participant nucleons, $<N_{part}>$, plays a crucial role in 
the presentation of PHOBOS or PHENIX centrality dependence data we first 
make a study on $<N_{part}>$. In the fixed target experiments the participant 
nucleons from the projectile nucleus with atomic number $A$, for instance, is 
estimated from
  \begin{equation}
  N_{part}^p=A*(1-\frac{E_{ZDC}}{E_{beam}^{kin}})
  \end{equation}
where $E_{ZDC}$ refers to the energy deposited in the Zero Degree Calorimeter
dominated by projectile spectator nucleons and $E_{beam}^{kin}$ is the kinetic 
energy of beam \cite{ahle}. However, in the collider experiments, in order to
obtain $<N_{part}>$ one has to relate the measurables to Monte Carlo 
simulations. In PHENIX, for instance, simulations for the response of the 
Beam-Beam Counter and the ZDC were used to calculate $<N_{part}>$ via a Glauber model \cite{phe1}. In PHOBOS $<N_{part}>$ is derived relating HIJING 
simulations to the signals in the paddle counter \cite{pho2}. 

\begin{figure}[ht]
\centerline{\hspace{-0.5in}
\epsfig{file=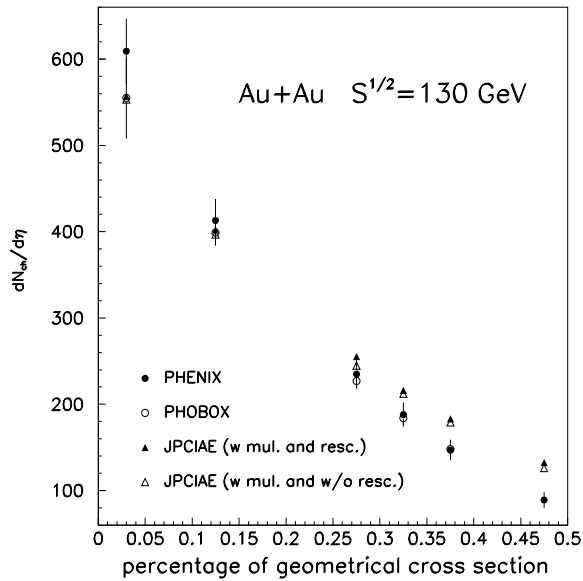,width=3.0in,height=3.0in,angle=0}}
\caption{The charged particle pseudorapidity density at mid-pseudorapidity 
 in $Au+Au$ collisions at $\sqrt s_{nn}$=130 GeV as a function of the 
 percentage of geometrical cross section.}
\label{fig3}
\end{figure}

 On the theory side, first, $<N_{part}>$ could be calculated geometrically 
(referred to as method a later) \cite{sa0}, number of participant nucleons 
from the projectile nucleus, for instance, reads
 \begin{equation}
  N_{part}^p(b)=\rho_0\displaystyle{\int} {dV\theta({R_p-[x^2+(b-y)^2+z^2]
                 ^{1/2}}})
                 \theta({R_T-(x^2+y^2)^{1/2}}).
 \end{equation}
Second, in the Glauber model $N_{part}$ is calculated by (referred to as 
method b later) \cite{esk1}   
 \begin{equation}
 N_{part}^p(b)=\int{d^2sT_A(\vec b-\vec s)[1-\exp(-\sigma_{in}T_B(\vec s))]}
              +\int{d^2sT_B(\vec s)[1-\exp(-\sigma_{in}T_A(\vec b-\vec s))]},
 \end{equation}
where $\sigma_{in}\approx$40 mb is the inelastic nn cross section at 
RHIC and $T_A$ refers to the nuclear thickness function of nucleus A.  
The third method is to count the participant or the spectator nucleons 
in the simulation for nuclear collisions and then to average over simulated 
events. However, there is multifarious in simulating programs and even in the 
definition of the participants and spectators. FRITIOF 7.02 (referred to as 
method d later) \cite{pi1} was popularly employed in the past. In FRITIOF the 
wounded nucleons, i.e., nucleons which suffer at least one inelastic collision, are counted and identified as $<N_{part}>$. It should be pointed out here that 
in FRITIOF leading nucleons undergo multiple scatterings and get excited 
(forming strings) during collisions, but produced particles from the string 
fragmentation do not have rescattering. Unlike JPCIAE, FRITIOF is not a 
transport model, there is no space-time coordinates associated with each 
particle. In JPCIAE simulations we have devised three counting methods: First, 
the leading nucleons involved in at least one inelastic nucleon-nucleon 
collision with string excitation are counted and identified as $<N_{part}>$. 
This is called method c. It should be mentioned that the JPCIAE results in 
Fig. 1 were calculated by $<N_{part}>$ from method c. Second, in method e, 
spectator nucleons are counted at the final state of JPCIAE simulations 
without rescattering (i.e., only nucleon-nucleon collisions are included), 
then $<N_{part}>$ is calculated through
 \begin{equation}
 <N_{part}>=(A+B)-<N_{spec}>,
 \end{equation}
where A and B refer to the atomic numbers of the target and projectile nuclei. 
Third, method f is the same as method e but JPCIAE simulations are 
with rescattering. The difference among method c, e, and f is that those 
nucleons which only experience two-body nucleon-nucleon collisions (without 
string formation) are included into $<N_{part}>$ in method e, while in method f 
even those nucleons which suffer collisions with other produced particles 
are also included into $<N_{part}>$. In other word, in method e and f 
the nucleons that are knocked out of the colliding nuclei by the produced 
particles are included into $<N_{part}>$. In emulsion chamber experiment, such 
nucleons (protons) are usually called `grey tracks'. 
 
 Fig. 2 (a) gives  $<N_{part}>$ calculated by different methods as a function 
of percentage of the geometrical (total) cross section and compares them 
with results of PHENIX (solid circles with error bar) \cite{phe1} and of PHOBOS 
(open circles with error bar) \cite{pho2}. The thin dotted, dashed, and solid 
curves in this panel are the results of method a (geometry), b (Glauber model, 
taken from \cite{esk1}), and c, respectively. Thick dotted, dashed, and solid 
curves are, respectively, the results of method d (FRITIOF), e, and f. One 
knows from Fig. 2 (a) that except method f $<N_{part}>$ from different methods 
are close to each other (the 10\%$-$15\% difference should contribute to the 
systematic error of the experimentally-extracted $<N_{part}>$ ) for most 
central collisions but the discrepancies among them increase with decreasing 
centrality in general. It is surprising that the results of geometry method are closest to the results of PHINEX or PHOBOS. In Fig. 2 (b) the ratios of 
$<N_{part}>$ from methods a, b, c, d, e and f to the corresponding results of 
PHENIX are given. 
       
 The charged particle pseudorapidity density at mid-pseudorapidity in $Au+Au$ 
collisions at $\sqrt s_{nn}$=130 GeV as a function of the percentage of 
geometrical cross section is given in Fig. 3. In this figure the full and 
open circles with error bar are the PHENIX \cite{phe1} and PHOBOS \cite{pho2} 
data, respectively. The full and open triangles, respectively, are the results 
of JPCIAE model with rescattering and without rescattering. The abscissa of the 
data point in this figure is set at the middle of the corresponding bin of   
percentage of the geometrical cross section. Globally speaking, the 
experimental data were reproduced fairly well by the JPCIAE model. However, 
the agreement between the experimental data and the JPCIAE model becomes less 
satisfied for peripheral collisions. One knows from this figure that the 
rescattering only leads to a few percent increase in charged multiplicity 
although rescattering might enhance yields of strangeness, $\Xi^- + 
\overline{\Xi^-}$ for instance, by a couple of times.  

\begin{figure}[ht]
\centerline{\hspace{-0.5in}
\epsfig{file=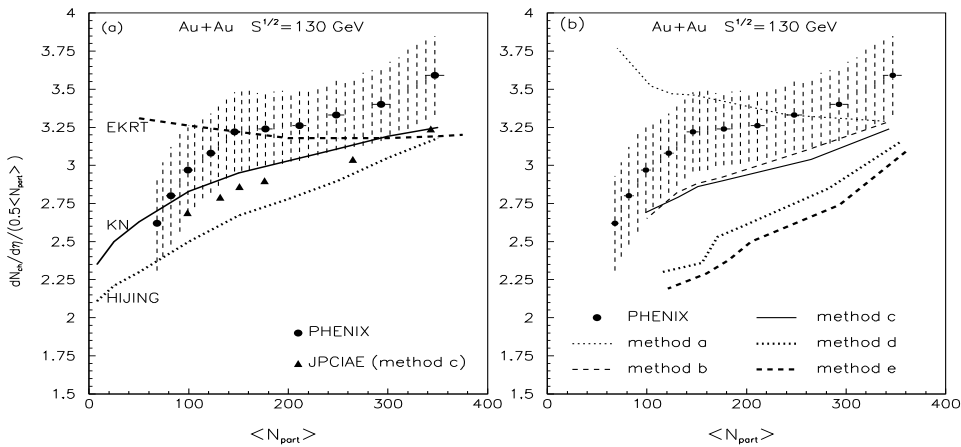,width=5.0in,height=2.3in,angle=0}}
\caption{The charged particle pseudorapidity density per participant pair  
 at mid-pseudorapidity in $Au+Au$ collisions at $\sqrt s_{nn}$=130 GeV as 
a function of the number of participant nucleons, $<N_{part}>$.}
\label{fig4}
\end{figure}

 In panel (a) of Fig. 4 we compare the PHENIX data of the charged particle 
mid-pseudorapidity density per participant pair (full circles with shaded area 
of systematic errors) \cite{phe1} with JPCIAE model (full triangles, 
$<N_{part}>$ from method c) and with HIJING model (dotted curve), KN model 
(solid curve), and EKRT model (dashed curve). One can see that except EKRT, 
all the other three models predict an increase of $\displaystyle 
(dn_{ch}/d\eta|_{\eta=0})/(0.5<N_{part}>)$ as a function of $<N_{part}>$ though
the theoretical results seem to underestimate the PHENIX data.
Fig. 4 (b) compares the PHENIX data to the JPCIAE results of $<N_{part}>$ 
calculated by method a (thin dotted curve), b (thin dashed curve), c (thin 
solid curve), d (thick dotted curve), and e (thick dashed curve), 
respectively. One sees from this panel that starting from a single result of 
charged particle mid-pseudorapidity density from the JPCIAE model but using 
$<N_{part}>$ from different methods, it is possible to lead $\displaystyle 
(dn_{ch}/d\eta|_{\eta=0})/(0.5<N_{part}>)$ to either increase or decrease with 
the increase of $<N_{part}>$. Although $<N_{part}>$ from method a are closest 
to the PHENIX results (cf. Fig. 2 (b)) the $(dn_{ch}/d\eta|_{\eta=0})/(0.5
<N_{part}>)$ from JPCIAE have actually centrality dependence 
opposite to the PHENIX result (cf. Fig.4 (b)) because the $\displaystyle 
(dn_{ch}/d\eta|_{\eta=0})$ from JPCIAE is higher than the PHENIX result for 
peripheral collisions (cf. Fig. 3). On the other hand, even though the 
discrepancy between $<N_{part}>$ from method c and PHENIX slightly increases 
with decrease of centrality (cf. Fig. 2 (b)) the $\displaystyle (dn_{ch}/d\eta|
_{\eta=0})/(0.5<N_{part}>)$ from JPCIAE with method c is close to the PHENIX 
data. If $\displaystyle dn_{ch}/d\eta|_{\eta=0}$ from EKRT model was not 
normalized by $<N_{part}>$ from method b, as did in \cite{esk1}, but by 
$<N_{part}>$ from method d the results of $\displaystyle (dn_{ch}/d\eta|_
{\eta=0})/(0.5<N_{part}>)$ might have somewhat similar centrality dependence 
as in PHENIX data. Therefore one learns here that it is hard using 
$\displaystyle (dn_{ch}/d\eta|_{\eta=0})/(0.5<N_{part}>)$ as a function of 
$<N_{part}>$ to distinguish various theoretical models for particle production 
since $<N_{part}>$ is not a well defined physical variable.  
   
In summary, we used a hadron and string cascade model, JPCIAE, to investigate 
the energy and centrality dependences of charged particle pseudorapidity 
density at mid-pseudorapidity in relativistic $p+\bar p$ and $A+A$ collisions. 
Both the relativistic $p+\bar p$ experimental data and the PHOBOS and PHENIX 
data of $Au+Au$ collisions at RHIC could be reproduced fairly well within the 
framework of the JPCIAE model without retuning any parameter. The JPCIAE model 
predictions for full RHIC energy $Au+Au$ collisions and for $Pb+Pb$ collisions 
at the ALICE energy are also given. This study shows that since $<N_{part}>$ 
is not a well defined physical variable both experimentally and theoretically 
it may be hard to use charged particle pseudorapidity density per participant 
pair at mid-pseudorapidity as a function of $<N_{part}>$ to distinguish various theoretical models for particle production.  
              
Finally, the financial supports from NSFC in China and DOE in USA are 
acknowledged.

\end{document}